\documentclass[letter]{aa}  

\usepackage{graphicx}
\usepackage{natbib}
\usepackage{txfonts}
\def\aj{AJ}                   
\def\apj{ApJ}                 
\def\aap{A\&A}                
\def\mnras{MNRAS}             
\def\pasa{PASA}               

\defcitealias{bonacic:07}{Paper~I}

\begin{document}

   \title{Stellar population synthesis of post-AGB stars: \\ the $s$-process in 
    MACHO\,47.2496.8}

   \author{A. Bona\v{c}i\'c Marinovi\'c
          \inst{1}
          \and
          M. Lugaro
          \inst{1,2}
          \and
	  M. Reyniers
          \inst{3}\fnmsep\thanks{Postdoctoral fellow of the Fund for Scientific Research, Flanders}
          \and
	  H. Van Winckel
          \inst{3}
          }

   \offprints{A. Bona\v{c}i\'c Marinovi\'c, bonacic@astro.uu.nl}

   \institute{Sterrekundig Instituut, Universiteit Utrecht, P.O. Box 80000,
         3508 TA Utrecht, The Netherlands\\
         \email{A.A.BonacicMarinovic@astro.uu.nl},
         \email{M.Lugaro@astro.uu.nl}
         \and
         Centre for Stellar and Planetary Astrophysics, School of Mathematical Sciences,
         Monash University, Victoria 3800, Australia
         \and
	 Instituut voor Sterrenkunde, Departement Natuurkunde en
         Sterrenkunde, K.U.Leuven, Celestijnenlaan 200D, 3001 Leuven,
         Belgium\\
         \email{maarten.reyniers@ster.kuleuven.be},
         \email{hans.vanwinckel@ster.kuleuven.be}
         }

   \date{Received -- May, 2007; accepted -- --, 2007}

   \authorrunning{Bona\v{c}i\'c Marinovi\'c et al.}
   \titlerunning{MACHO\,47.2496.8}

\abstract
{The low-metallicity RV\,Tauri star MACHO\,47.2496.8, recently discovered in the Large 
Magellanic Cloud, is highly enriched in carbon and heavy elements produced by the 
$slow$ neutron capture process ($s$-process), and is most probably a genuine 
post-C(N-type) asymptotic giant branch (AGB) star.  The intrinsic
  interpretation of the enrichement is further strengthened by
  detection of a significant infrared excess. The circumstellar dust is the relic of a
  recent episode of heavy mass loss.  We use 
the analysis of the abundances of MACHO\,47.2496.8 to constrain free parameters in AGB models.}
{We test which values of the free parameters describing uncertain physical mechanisms 
in AGB stars, namely the third dredge-up and the features of the $^{13}$C neutron 
source, produce models that better match the abundances observed in 
MACHO\,47.2496.8.}
{We carry out stellar population synthesis coupled with $s$-process
nucleosynthesis using a synthetic
stellar evolution code.}
{The $s$-process ratios observed in MACHO\,47.2496.8 can be matched by 
the same models that explain the $s$-process ratios of Galactic AGB and post-AGB stars 
of metallicity $> Z_{\odot}$/10, except for the choice of the effectiveness of 
$^{13}$C as a neutron source, which has to be lower by roughly a factor of 3 to 6. 
The less effective neutron source for lower metallicities
is also required when comparing population synthesis results to 
observations of Galactic halo $s$-enhanced stars, such as Pb stars. The 
$^{12}$C/$^{13}$C ratio in MACHO\,47.2496.8 cannot be matched
simultaneously and requires the occurrence of extra-mixing 
processes. 
}
{The confirmed trend of the decreased efficiency of the $^{13}$C neutron source with 
metallicity requires an explanation from AGB $s$-process models. The 
present work is to date the first comparison between theoretical models and the 
detailed abundances of an extragalactic post-AGB star.}


   \maketitle

\section{Introduction}\label{sect:intro}

{\em Slow} neutron capture ($s$-process) elements are synthesized in the intershell 
region between the C$-$O core and the convective H-rich envelope of asymptotic giant 
branch (AGB) stars. The main neutron source is believed to be $^{13}$C nuclei, 
releasing neutrons via the ${\rm ^{13}C}(\alpha,n)^{16}{\rm O}$ reaction in a thin 
layer of the intershell (the $^{13}$C {\it pocket}, see e.g. 
\citealp{gallino:98} for details). The 
$s$-process elements are mixed to the stellar surface by recurrent episodes of 
third dredge-up (TDU), where they are observed. A high abundance of $^{13}$C in the 
pocket can be 
produced by $^{12}$C+$p$ reactions if protons are mixed from the convective envelope 
into the radiative $^{12}$C-rich intershell. This mixing would likely occur at the 
end of each TDU episode, when a sharp discontinuity is left between the convective 
and the radiative regions. However, it is not yet clear what mechanism 
produces this mixing. The extent of the mixing, which makes the effectiveness of the $^{13}$C neutron 
source, is the most uncertain parameter in $s$-process models \citep[see 
e.g.][for discussion]{busso:99,lugaro:03,herwig:05}.

Post-AGB stars are in the fast evolutionary phase between the AGB
and white dwarf tracks. The strong mass loss at the end of the AGB has
stopped, but they are not yet hot enough to ionise their circumstellar
medium and to produce a planetary nebula. Their elemental abundances are the result of 
their evolutionary history, thus they can be used to probe and constrain the
nucleosynthesis that takes place in AGB stars, in particular
the $s$-process.

A peculiar post-AGB object, MACHO\,47.2496.8, was found in the Large
Magellanic Cloud (LMC) by \citet{pollard:00} and analysed in detail by
\citet{reyniers:07}. Its main features are 
a very low metallicity ([Fe/H]=$-$1.4), excess of carbon over oxygen (C/O$>$2 with  
$^{12}$C/$^{13}$C=200$\pm$25),
an enhancement of 1.2 dex of light $s$-process elements (ls) compared to iron,
a strong enhancement of heavy $s$-process elements (hs) compared to iron of 2.1 dex,
and a similar enhancement of lead over iron.
In \citet{reyniers:07}, it was argued that the luminosity
($\sim$5000\,L$_{\sun}$) and the specific pulsational behaviour (RV\,Tauri like)
of MACHO\,47.2496.8
favour an {\em intrinsic} origin of the $s$-process enrichement, although
an extrinsic scenario in which the enrichment is caused by a former AGB
companion, now on the white dwarf track, could not be excluded.

\begin{figure}
   \centering
   \includegraphics[width=0.45\textwidth]{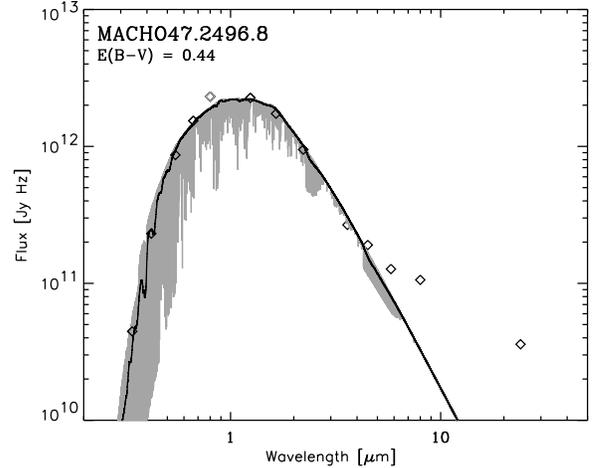}
\caption{The spectral energy distribution of MACHO\,47.2496.8 as in
  \citet{reyniers:07}, but updated with Spitzer SAGE fluxes, revealing
  the presence of circumstellar dust. Diamonds are the measured
  magnitudes (from blue to red): Geneva U, B, V and
  Cousins R taken with C2+Euler; I from DENIS (in gray), SAAO J, H, K, 
  and the Spitzer SAGE fluxes (3.6, 4.5, 5.8, 8.0, and 24\,$\mu$m). The MARCS
  model is shown in gray, while a smoothed version is shown with a full
  black line. The I magnitude could not be fitted, possibly due to a
  phase difference.}
      \label{fig:sedspitzer}
   \end{figure}

We have updated the Spectral Energy Distribution (SED) of MACHO\,47.2496.8
presented in \citet{reyniers:07} by adding the newly released Spitzer SAGE data
\citep{meixner:06} of the source. The new data reveal a small but clear infrared
excess starting around 5\,$\mu$m (Fig.~\ref{fig:sedspitzer}).  
In the Galaxy, the RV\,Tauri pulsators with a dust excess are
dominated by binaries, consisting of a post-AGB primary and an
unevolved companion. In these systems, the dust is trapped in a stable
circumbinary disc
\citep{deruyter:06}. The LMC RV\,Tauri stars detected by the Macho
experiment are also dominated by disc sources \citep{reyniers:07b}
and the SAGE data reveal that the typical colours yield a K$-$[5.8$\mu$m] $>$
2.5 and [8$\mu$m]$-$[24$\mu$m] indicative of a dust excess with a hot
dust component. 

The colours
of MACHO\,47.2496.8 (K$-$[5.8]=0.8 and [8]$-$[24]=2.4) are significantly
different, showing that the excess is much colder than in the suspected disc sources. We
conclude that the dust excess of MACHO\,47.2496.8 corroborates the
post-AGB status of the object in which the excess is a relic of a
recent phase of dusty mass loss.  The object is an ideal source to
study the dust formation in metal poor conditions.
In principle, it could still represent the post-AGB phase of a
former, extrinsically enriched, CH star 
\citep[about 5\% of the carbon stars in the][catalog are extrinsically
$s$-process and carbon enriched CH-stars]{stephenson:89} 
but we consider that it is much more likely that the object is a genuine 
intrinsically enriched post-AGB star.

In this contribution, we focus on the modelling of the chemical
content assuming that the object is indeed intrinsically enriched. 
By comparing recent stellar population calculations
including nucleosynthesis of $s$-process elements to
observations of Galactic stars,
\citet[hereafter Paper~I]{bonacic:07} have set
constraints on several free parameters included in their models:
the minimum core mass for TDU, the TDU efficiency (${\rm \lambda}$), the effectiveness of 
the $^{13}$C neutron source ($^{13}$C$_{\rm eff}$), and the size in mass of the 
$^{13}$C pocket. In this paper we use those models
to interpret the abundances of MACHO\,47.2496.8 and determine whether 
they can be reproduced using the same set of values of the free parameters
found in ${\rm Paper~I}$. 


\section{Stellar population synthesis models}\label{sect:models}
   \begin{figure*}
   \centering
   \includegraphics[width=\textwidth]{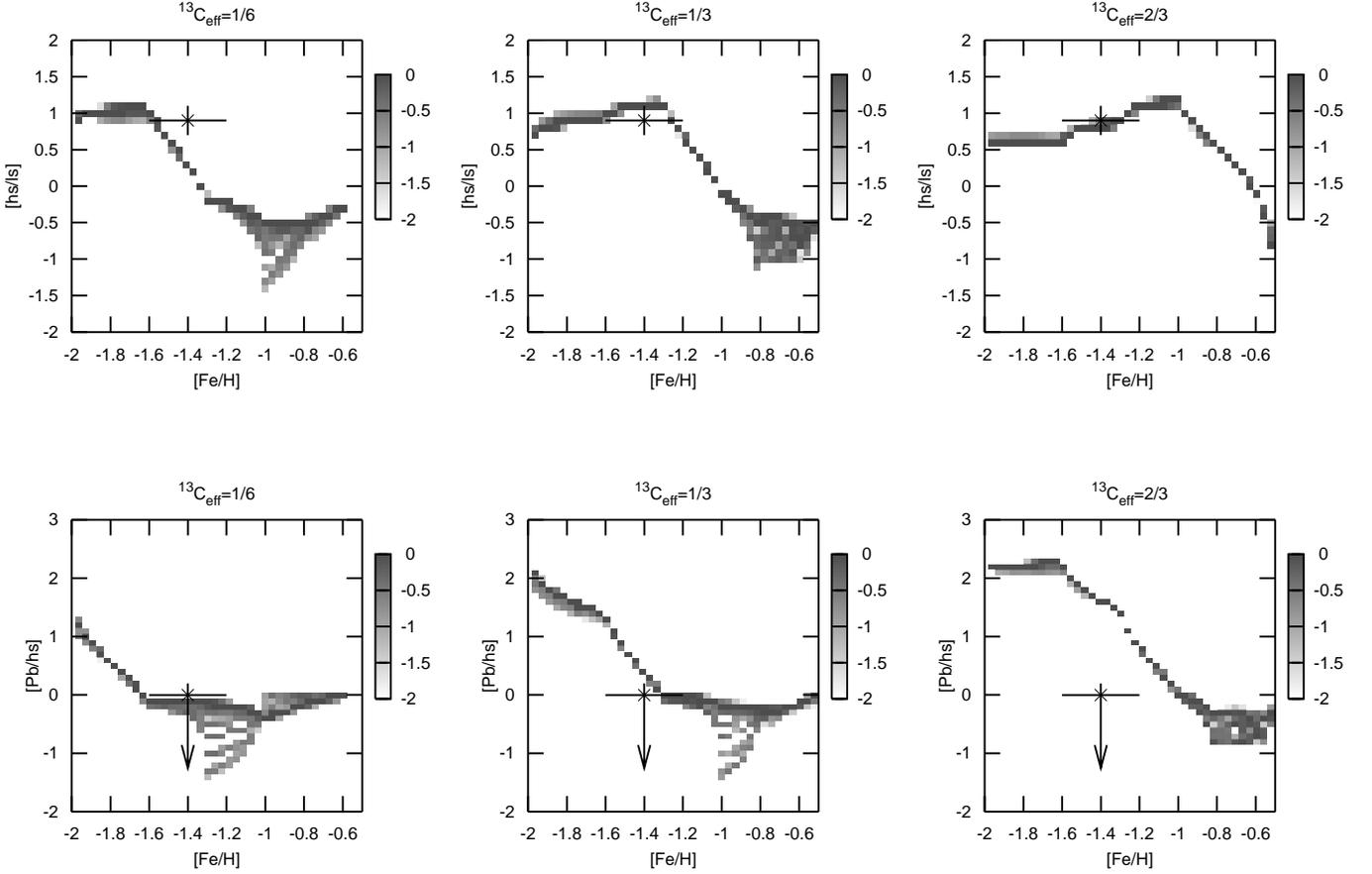}
      \caption{Distribution of [hs/ls] (upper panels) and [Pb/hs]
        (lower panels) ratios in synthesized populations of
        intrinsically $s$-process enhanced post-AGB stars,
        calculated with different values of $^{13}{\rm C_{eff}}$,
        as indicated over each panel.
        The grey scale is a logarithmic measure of the number
        distribution of stars over the plotted $s$-process index. The
        number density is weighed by the initial mass function and by
        the time each star spends in an abundance bin, and then
        normalized for each metallicity.
        The results are compared to the $s$-process element
        ratios observed in MACHO\,47.2496.8, indicated by the crosses.
        The size of the vertical and horizontal lines indicate the
        observational errors of the data and the arrow indicates
        that the observed [Pb/hs] ratio is an upper limit.
        }
      \label{fig:s-ratios}
   \end{figure*}

We calculated populations of post-AGB stars with our rapid synthetic
stellar evolution code, which includes $s$-process
nucleosynthesis based on the models of
\citet{gallino:98}. This code is described in detail
in ${\rm Paper~I}$ and the population synthesis procedure is done
in the same fashion as in ${\rm Paper~I}$. 

We run our models on a grid of 50 metallicity values, [Fe/H], linearly
separated, in the range $-2.0<{\rm [Fe/H]}<-0.5$ and 500 initial masses,
$M_{\rm i}$, logarithmically separated, in the range
$0.7~{\rm M_{\odot}}<M_{\rm i}<1.2~{\rm M_{\odot}}$,
weighed by the initial mass function of \citet{1993MNRAS.262..545K}.
In contrast to ${\rm Paper~I}$, we do not consider an age-metallicity
relation to calculate the range of masses given that
MACHO\,47.2496.8 belongs to the LMC and age-metallicity relations for
the LMC are uncertain due to the low precision of age estimates
\citep[see e.g.][]{cole:05}. In addition, if we apply the
age-metallicity relation from \citet{cole:05} 
the most massive AGB objects obtained with the metallicity of
MACHO\,47.2496.8 are not massive enough to experience
dredge-up (about 0.85 M$_{\odot}$ in their zero age main sequence).

The synthetic models are carried out applying the
prescription of \citet{1993ApJ...413..641V} to account for the mass
loss. We employ the free-parameter values found in ${\rm Paper~I}$,
which provide the best match for the overall properties of Galactic
$s$-enhanced AGB and post-AGB stars: a shift of the minimum
core mass for TDU $\Delta M_{\rm c}^{\rm min}=-0.065$M$_{\odot}$ with respect to the 
models of \citet{2002PASA...19..515K}, a minimum asymptotic TDU efficiency 
${\rm \lambda}_{\rm min}=0.2$, and a $^{13}$C-pocket size, given as a fraction of the  
intershell mass $f_{\rm ^{13}C,IS}=1/40$. In particular, the relatively high
${\rm \lambda}_{\rm min}$ is needed in order to match the number of $s$-process
enhanced Galactic post-AGB star that are also carbon rich \citepalias[see][]{bonacic:07}.
Once the TDU parameters are fixed, observed $s$-process enhancements such
as the [Zr/Fe] ratio in Galactic post-AGB stars, can be matched by
adjusting $f_{\rm ^{13}C,IS}$ to a somewhat smaller value than that usually 
employed in single star models with less efficient TDU \citep{gallino:98,goriely:00}.
With these choices of the free parameters the abundances observed in all the
other types of intrinsic AGB $s$-enhanced stars (namely MS, S, SC and C stars) are 
also reproduced \citepalias[see][]{bonacic:07}.

The parameters described above affect the overall enhancement of $s$-process elements 
with respect to Fe at the stellar surface, i.e., [ls/Fe] and [hs/Fe]. However, they 
do not affect the relative distribution of $s$-process abundances, represented by the 
[hs/ls] and [Pb/hs] ratios, which is mainly a function of $^{13}{\rm C}_{\rm eff}$. 
In ${\rm Paper~I}$ we found that, for Galactic objects with ${\rm [Fe/H]}\gtrsim-1$, 
$^{13}{\rm C}_{\rm eff}$ ranges between approximately $2/3$ and $4/3$ of the standard 
value introduced by \citet[see details in Paper~I]{gallino:98}, while a value reduced by
roughly a factor of 6 to 12 is needed to fit Pb stars, which are lower metallicity,
extrinsically enriched halo objects. Here we compare our results with a range of values
for this parameter to the observational data of MACHO\,47.2496.8.

\section{Results}

We select $s$-process enhanced post-AGB stars from our models by
choosing those TP-AGB objects that have ${\rm [ls/Fe]}\geq0.25$ or
${\rm [hs/Fe]}\geq0.25$ and an envelope mass $\leq$ 0.03 M$_{\odot}$.
Figure~\ref{fig:s-ratios} shows stellar population results
computed with different ${\rm ^{13}C}_{\rm eff}$ values and
compared to the observed $s$-process element ratios of
MACHO\,47.2496.8 measured by \citet{reyniers:07}.
The grey scale measures the distribution of stars over the
plotted $s$-process ratio in terms of metallicity.
The darker area represents the contribution of stars with
initial mass $M_{\rm i}\approx\, $M$_{\odot}$, which are the most numerous
according to the initial mass function that we consider.
A good match to the observations for both 
[hs/ls] and [Pb/hs] ratios is obtained with $^{13}{\rm C_{eff}}$ $1/3$ to $1/6$ of the standard 
value. Any smaller $^{13}{\rm C_{eff}}$
value is inconsistent given that the pattern of 
[hs/ls] as a function of metallicity would
shift towards lower metallicities
\citepalias[for a detailed description see][]{bonacic:07}.

The models reported in Fig.~\ref{fig:s-ratios} can also be employed to interpret 
the composition of the Galactic post-AGB star IRAS07134+1005. This object 
has metallicity [Fe/H]$\sim-1$ and a very high heliocentric
velocity \citep{2000A&A...354..135V}, suggesting
that it belongs to the Galactic halo rather than the disc. In 
${\rm Paper~I}$
it was discussed that this object ``apparently needs a somewhat smaller 
$^{13}{\rm C}_{\rm eff}$'' than Galactic post-AGBs of higher metallicity; however, 
because of the age-metallicity relation employed in ${\rm Paper~I}$ 
it was not possible to model post-AGB stars of such low metallicity. With the 
models presented here, we can now confirm that the [hs/ls] ratios approximately 
equal to zero observed in IRAS07134+1005 can be well matched by $^{13}{\rm C}_{\rm 
eff}$=1/3.
The same conclusion holds for the intrinsic halo C star HD\,189711 with [Fe/H]=$-1.14$ and
[hs/ls]=0.7 shown in Fig. 4 of Paper I.
We also note that this value of $^{13}{\rm C}_{\rm eff}$ is not in disagreement with 
observations of different types of extrinsic $s$-process stars (such as halo CH 
giant, 
halo yellow symbiotic, and Pb stars) in the same metallicity range 
\citepalias[see Fig.~10 of][]{bonacic:07}.

   \begin{figure}
   \centering
   \includegraphics[width=0.5\textwidth]{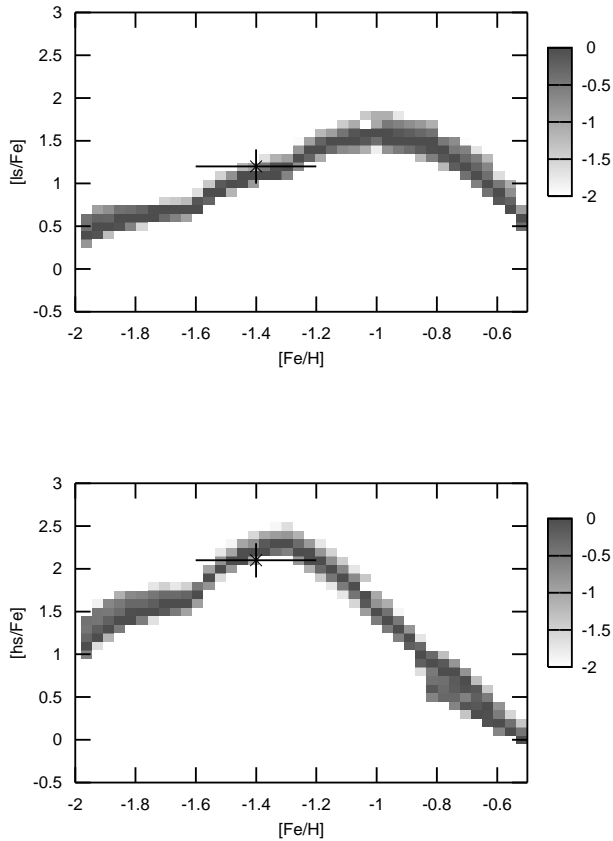}
      \caption{Distribution of [ls/Fe] (upper panel) and [hs/Fe]
        (lower panel) ratios in a synthetic population of
        intrinsically $s$-process enhanced post-AGB stars,
        calculated by assuming $^{13}{\rm C_{eff}}=1/3$.
        The grey scale is a logarithmic measure of the number
        distribution of stars over the plotted $s$-process index. The
        number density is weighed as described in the caption
        of Fig~.\ref{fig:s-ratios}
        The results are compared to [ls/Fe] and [hs/Fe] ratios
        measured in MACHO\,47.2496.8, indicated by the crosses.
        The size of the vertical and horizontal lines indicate the
        errors on the data measurements.
        }
      \label{fig:s_Fe}
   \end{figure}

The [hs/Fe] and [ls/Fe] ratios of MACHO\,47.2496.8 are well reproduced by our results for 
intrinsically $s$-process enhanced post-AGB stars with the choice of $^{13}{\rm C}_{\rm eff}=1/3$ 
to 1/6 (the choice of 1/3 is represented in Fig.~\ref{fig:s_Fe}) and without the need for any 
modification of the remaining free parameters, which, as described above, were already set 
according to the properties of Galactic post-AGB stars.

A problem arises when considering the C/O and, in particular, the $^{12}$C/$^{13}$C ratio in
MACHO\,47.2496.8. 
With the high TDU assumed in our model we obtain 
C/O$>$10 and $^{12}$C/$^{13}$C ratios at least an order of magnitude higher than those 
observed. A better solution is possible by decreasing ${\rm \lambda}_{\rm min}$. For example, 
using ${\rm \lambda}_{\rm min}$=0.1 we obtain C/O$>$4 and $^{12}$C/$^{13}$C$>$300. Then, we 
need to increase $f_{\rm ^{13}C,IS}$ to 1/20 to match the [ls/Fe] and [hs/Fe] ratios. 
The problem is that, with 
this choice, we do not match the number of $s$-process enhanced Galactic post-AGB 
star that are also carbon rich \citepalias[see][]{bonacic:07} and it is difficult to 
find a consistent solution for these two different constraints. A more promising 
explanation can be found by remembering that model predictions always produce too 
high $^{12}$C/$^{13}$C ratios with respect to observations, both for red giant and 
AGB stars. 
Extra-mixing processes, also sometimes called ``deep mixing'' or cool bottom 
processing, that would enable envelope material to suffer proton captures, thus 
transforming $^{12}$C into $^{13}$C, have been invoked to explain, e.g., the low 
$^{12}$C/$^{13}$C ratios observed in giant stars \citep{gilroy:89} and 
in carbon stars \citep{2001ApJ...559.1117A}, as well as in meteoritic silicon 
carbide grains from AGB stars \citep[e.g.][]{zinner:06}. We cannot rule 
out that such processes could also have affected the C composition of 
MACHO\,47.2496.8. Recently, mixing instabilities have been found to occur in 
first giant branch stars because 
of a small inversion in the molecular gradient, just above the H-burning shell, 
where the $^{3}$He($^{3}$He,2$p$)$^{4}$He reaction is activated 
\citep{eggleton:06,charbonnel:07}. This physical mechanism likely leads to 
modifications of the CNO abundances, as observed, and also would be at work during 
the AGB phase (M. Cantiello, personal communication). Note that extra mixing on the red 
giant branch would not be enough to explain the $^{12}$C/$^{13}$C ratio 
observed in MACHO\,47.2496.8.

\section{Conclusion}\label{sect:maincnclsns}

We have compared spectroscopic observations of the post-AGB star MACHO\,47.2496.8 
in the LMC to results obtained by carrying out stellar 
population synthesis coupled with $s$-process nucleosynthesis in order to obtain 
constraints on the physics of AGB stars.  
The result is that the values of the free parameters needed to match 
MACHO\,47.2496.8 are consistent with the values we found 
in ${\rm Paper~I}$ for Galactic $s$-process enhanced stars:
\begin{enumerate}
\item $\Delta M_{\rm c}^{\rm min}= -0.065$M$_{\odot}$,
\item ${\rm \lambda}_{\rm min}=0.2$, and
\item $f_{\rm ^{13}C,IS}=1/40$.
\item Objects with metallicity ${\rm [Fe/H]}\lesssim-1$ are generally well
reproduced by a lower value of $^{13}{\rm C}_{\rm eff}$ than objects with 
metallicity ${\rm [Fe/H]}\gtrsim-1$.
\end{enumerate}
The strong agreement among results for different populations and for stars in 
different galaxies reinforces our confidence in the findings of ${\rm Paper~I}$. 
Regarding point (4) above, for IRAS07134+1005 ([Fe/H]=$-$1) and 
HD\,189711 ([Fe/H]=$-$1.15) we 
have to lower $^{13}{\rm C}_{\rm eff}$ by roughly a factor of 3, with respect to 
higher-metallicity objects, for 
MACHO\,47.2496.8 ([Fe/H]=$-$1.4) we have to lower $^{13}{\rm C}_{\rm eff}$ by 
roughly a factor of 3 to 6, while for Pb 
stars ([Fe/H]$\lesssim -2$) we have to lower $^{13}{\rm C}_{\rm eff}$ by roughly a 
factor of 6 to 12. This trend requires for an explanation within AGB $s$-process 
models. 
 We conclude that the heavy element chemical composition of the metal-poor LMC
  star MACHO\,47.2496.8 is consistent with model predictions of an
  intrinsic $s$-process enhanced star, but that an extra-mixing process
  is needed to account for the carbon isotopic and C/O ratios.
Future work will have to address the question of whether data from the still small but 
expanding set of observed extragalactic AGB stars \citep[see e.g.][]{delaverny:06} can be 
fitted with our current choice of model parameters. 

\begin{acknowledgements}
ML gratefully acknowledges the support of NWO through the VENI grant; MR acknowledges
financial support from the Fund for Scientific Research - Flanders (Belgium).
\end{acknowledgements}

\end{document}